\documentclass[a4paper,english,aps,floats,onecolumn,showpacs,nofootinbib]{revtex4}
\usepackage{pslatex}
\usepackage[T1]{fontenc}
\usepackage{graphicx}
\usepackage{epsfig}

\usepackage{calc}
\usepackage{ifthen}
\usepackage{subfigure}

\newcommand{\nc}{\newcommand}

\newcommand{\bea}{\begin{eqnarray}}
\newcommand{\eea}{\end{eqnarray}}

\nc{\hf}{\frac{1}{2}}

\newcommand{\tdN}{\tilde{N}}

\nc{\renc}{\renewcommand}
\nc{\eqs}[2]{\mbox{Eqs.~(\ref{#1},\,\ref{#2})}}
\nc{\eq}[1]{\mbox{Eq.~(\ref{#1})}}
\nc{\figs}[2]{\mbox{Figs.~(\ref{#1},\,\ref{#2})}}
\nc{\fig}[1]{\mbox{Fig~.(\ref{#1})}}
\nc{\be}[1]{\begin{equation} \mbox{$\label{#1}$}}
\nc{\ee}{\vspace{0.1cm}\end{equation}}

\newcommand{\bean}{\begin{eqnarray*}}
\newcommand{\eean}{\end{eqnarray*}}

\begin{document}
\title{A Unitarity-Conserving Higgs Inflation Model}
\author{Rose N. Lerner}
\email{r.lerner@lancaster.ac.uk}
\author{John McDonald}
\email{j.mcdonald@lancaster.ac.uk}
\affiliation{Cosmology and Astroparticle Physics Group, University of
Lancaster, Lancaster LA1 4YB, UK}

\begin{abstract}

    Scalar field models of inflation based on a large non-minimal coupling to gravity $\xi$, in particular Higgs Inflation, may violate unitarity at an energy scale $\Lambda \sim M_{p}/\xi \ll M_{p}$. In this case the model is incomplete at energy scales relevant to inflation. Here we propose a new unitarity-conserving model of Higgs Inflation. The completion of the theory is achieved via additional interactions which are proportional to products of the derivatives of the Higgs doublet. The resulting model differs from the original version of Higgs Inflation in its prediction for the spectral index, with a classical value $n = 0.974$. In the case of a non-supersymmetric model, quantum corrections are likely to strongly modify the tree-level potential, suggesting that supersymmetry or a gauge singlet scalar inflaton is necessary for a completely successful model.

\end{abstract}

\maketitle

\section{Introduction}

        There has been much interest recently in models of inflation using scalar fields non-minimally coupled to gravity, originally proposed in \cite{salopek}. (See also \cite{early}.)  This was primarily motivated by the idea of using the Standard Model Higgs as the inflaton (`Higgs Inflation') \cite{bs1}. Variants include Higgs Inflation with scalar dark matter \cite{clark}, `S-inflation' due to a dark matter scalar coupled to the Standard Model \cite{rl1}, a supersymmetric version of Higgs Inflation \cite{tj} and an extension to include neutrino masses \cite{shafi}. Generalization of the non-minimal coupling to gravity was discussed in \cite{park}\footnote{A alternative approach to Higgs Inflation, based on derivatives of the Higgs coupled to gravity, was presented in \cite{germani}.}.

      However, the naturalness of these models has been questioned, specifically whether or not unitarity is violated in Higgs scattering mediated by graviton exchange at a scale $\Lambda \sim M_p / \xi \ll M_p$. Here $\xi$ is the value of the non-minimal coupling, which must be of order $10^4$ in order to account for the observed density perturbation.
In particular, in \cite{barbon} it was noted that the effective coupling in tree-level graviton-mediated Higgs scattering becomes strong at $E \sim \Lambda$, while in \cite{b1} it was concluded that unitarity would be violated in graviton-mediated Higgs scattering at $E \sim \Lambda$.

   These analyses were based on the original Higgs Inflation model, which considered a single real Higgs scalar in the unitary gauge and neglected gauge interactions. In \cite{nat} it was noted that there are no strong coupling or unitarity-violating interactions in the single scalar model when considered in the Einstein frame, indicating that
the apparent strong coupling or unitarity-violating effects in the Jordan frame at $E \sim \Lambda$ do not occur. This can be understood in terms of a cancellation of the leading s-, t- and u-channel contributions to the graviton-mediated Higgs amplitude in the Jordan frame \cite{cancel,hertz}. However, once longitudinal gauge fields are included in the unitary gauge (or, equivalently, Goldstone bosons in a covariant gauge), the Jordan frame cancellation of the graviton-mediated Higgs scattering amplitude no longer occurs \cite{hertz,b2}. This manifests itself in the Einstein frame as non-renormalizable interactions which cannot be eliminated by field redefinitions.

        However, while unitarity is violated in tree-level scattering, it was shown in \cite{hw} that perturbation theory will break down before the energy of unitary violation is reached. Specifically, for the case of s-channel scattering mediated by graviton exchange, it was shown that the imaginary part of the 1-loop contribution to the amplitude is half of the tree-level contribution at the energy of unitarity-violation. As noted in \cite{nat}, this leads to the possibility that strong-coupling itself is the "new physics" required to maintain unitarity. This is supported by the observation of \cite{hw} that, in the large-$N$ limit (where $N$ is roughly the number of particles contributing to the loop corrections), the all-order graviton-mediated scattering cross-section (excluding graviton loops) is unitary at all energies, even though the tree-level cross-section violates unitarity. The possibility that strong coupling could ensure unitarity-conservation was noted earlier in \cite{bs2}. The essential point is that if strong coupling can deal with the apparent unitarity violation in particle scattering processes, then
the action of the theory is complete as is, requiring no new terms. The effective potential and the analysis of inflation can then be carried out by calculating with this action in the conventional way \cite{hi1}.

    Logically, the action of the original Higgs Inflation model is either consistent or inconsistent as a quantum field theory. If it is an inconsistent theory then we expect unitarity to be violated at some energy, requiring a completion of the theory. However, if the theory is consistent, then we would expect any calculation which appears to violate unitarity to be modified as the energy approaches that of unitarity violation. This appears to be the case in Higgs Inflation, with higher-order corrections to the scattering amplitude becoming important as the energy approaches that at which tree-level unitarity is violated. Therefore Higgs Inflation has the qualitative behaviour of a consistent theory. However, since a non-perturbative analysis is necessary in order to establish unitarity conservation in Higgs Inflation, it may be difficult to either prove or disprove unitarity conservation. In this case the best strategy would be to consider both possibilities and use collider experiments and precision CMB observations to establish whether Higgs Inflation is consistent with observations. This strategy is feasible because of the uniquely predictive nature of Higgs Inflation. The inflation observables, in particular the spectral index, are entirely determined by Standard Model couplings. Therefore precision measurement of the spectral index and the Higgs mass $m_{H}$ can, in principle, allow the nature of Higgs Inflation to be determined experimentally.

          The case where unitarity is conserved in Higgs Inflation has been extensively studied in \cite{bs2,hi1}, where the RG-improved effective potential was calculated and the spectral index as a function of Higgs mass determined. In this paper we consider the alternative case where unitarity is violated at $E \sim \Lambda$. In this case we must add new terms to the action to restore unitarity. The concern expressed in \cite{barbon,b1} is that such
new terms necessarily include Higgs potential terms suppressed by powers of $\Lambda$, spoiling the flatness of the potential and ruling out slow-roll inflation. However, this is an assumption. Our goal here is to derive the minimal modification of Higgs Inflation necessary to restore unitarity and to show that it can, in principle, support successful inflation.

          In Section 2 we review tree-level unitarity violation in the original Higgs Inflation model. In Section 3 we introduce a new unitarity-conserving Higgs Inflation model. In Section 4 we discuss the cosmology of this model, showing that it makes a quite different prediction for the spectral index from the original Higgs Inflation model. In Section 5 we present our conclusions.

\section{Tree-level unitarity violation in Higgs Inflation}

       We first consider tree-level unitarity violation due to graviton-mediated Higgs scattering in Higgs Inflation.  In the Jordan frame the action for Higgs Inflation (including the Higgs doublet and gauge fields) is
\be{e0}
 S_J =  \int d^4 x \; \sqrt{-\!g} \; \left( - \frac{M^2R}{2} - \xi H^{\dagger}H R  +  g^{\mu\nu}\left(D_\mu H\right)^{\dagger}\left(D_\nu H\right)    -\frac{1}{4} F_{\mu\nu} F^{\mu\nu} - V(|H|) \right)
~,\ee
where
\be{e1}
\label{Jpot}
V(|H|)  = \lambda \left(\left(H^{\dagger}H\right) - \frac{v^2}{2}\right)^2  ~.\ee
(In \eq{e0} the gauge kinetic term represents the kinetic terms for all gauge fields.) In the following we will set $M = M_p$, as the Higgs vacuum expectation value is negligibly small compared with $M_{p}$. The Einstein frame action is obtained by first performing a
conformal rescaling of the metric
\be{2} \tilde{g}_{\mu\nu} = \Omega ^2 g_{\mu\nu} ~,\ee
where
\be{3}\label{omegaeq} \Omega ^2 = 1 + \frac{2 \xi H^{\dagger}H}{M_p^2} ~.\ee
In terms of this metric the action becomes
\be{e4}  S_{E} = \int d^{4} x  \sqrt{-\tilde{g}} \left( - \frac{M_{p}^{2}}{2}\tilde{R} + \frac{1}{\Omega^{2}} \tilde{g}^{\mu\nu}\left(D_\mu H\right)^{\dagger}\left(D_\nu H\right)  + \frac{3\xi^2}{\Omega^4 M_p^2}\tilde{g}^{\mu\nu} \partial_\mu\left(H^{\dagger} H\right) \partial_\nu\left(H^{\dagger} H\right) -\frac{1}{4} F_{\mu\nu} F^{\mu\nu} - \frac{V(|H|)}{\Omega^4} \right) ~,\ee
where $\tilde{R}$ is the Ricci scalar with respect to $\tilde{g}_{\mu \nu}$ and indices are raised with $\tilde{g}^{\mu \nu}$.

   Tree-level unitarity violation due to graviton-mediated Higgs scattering in the Jordan frame manifests itself in the Einstein frame via the non-minimal kinetic terms for $H$ from the second and third terms in \eq{e4}. The simplest way to consider unitarity violation is to consider the $\langle H \rangle \rightarrow 0$ limit, where the physical degrees of freedom are the four real scalars of $H$ and the
transverse gauge degrees of freedom. In the case with a single real scalar ($H \rightarrow h/\sqrt{2}$) and no gauge fields, as originally considered in Higgs Inflation, the non-minimal kinetic term can be eliminated by a redefinition of $h$ to $\chi$ via
 \be{e5} \frac{d\chi}{dh} = \sqrt{\frac{\Omega ^2 + 6 \xi^2h^2/M_P^2}{\Omega ^4}} ~.\ee
Then
\be{e6}
S_E = \int d^4x\sqrt{-\tilde{g}}\Big( - \frac{M_p^2\tilde{R}}{2} + \frac{1}{2}\partial _\mu \chi \partial^{\mu} \chi - U(\chi)\Big)
~,\ee
where $U(\chi) = V(h)/\Omega^4$. In this case there are no interactions\footnote{One concern is that the non-polynomial potential is difficult to handle as a quantum field theory. However, we believe this is a quite different issue from tree-level unitarity violation associated with the non-minimal coupling to gravity in the Jordan frame. Since tree-level unitarity violation in $2 \rightarrow 2$ Higgs scattering via graviton-exchange is independent of the potential, the analogous interactions in the Einstein frame should also be independent of the potential. We will comment further the issue of the non-polynomial potential in our conclusions.} which lead to tree-level unitarity violation in
$\chi$-$\chi$ scattering, which is equivalent to $h$-$h$ scattering since $\Omega \approx 1$ in the vacuum. The absence of
unitarity-violating interactions in the Einstein frame at $E \sim \Lambda$ is consistent with the cancellation of the leading s-, t- and u-channel contributions to graviton-mediated Higgs scattering in the Jordan frame \cite{cancel,hertz}.

However, with more than one scalar, it is no longer possible to redefine the scalar fields to have canonical kinetic terms, since this would require the non-minimal kinetic term for the field $\phi_{i}$ to be a function of $\phi_{i}$ only. As a result, there are Einstein frame interactions such as
\be{e8}   \frac{3 \xi^{2}\phi_{i}\phi_{j}}{\Omega^{4} M_{p}^{2}}\partial_{\mu}\phi_{i} \partial^{\mu} \phi_{j}      ~,\ee
where $\phi_{i}$ ($i = 1,...4$) are the 4 real scalars in $H$.
These interactions lead to a tree-level scattering amplitude for $\phi_{i} \phi_{i} \rightarrow \phi_{j} \phi_{j}$ which is of the order of $(E/\Lambda)^2$. The corresponding cross-section will therefore violate unitarity at $E \gtrsim \Lambda$. 
The same result may also be obtained in the unitary gauge with $\langle H \rangle = v$, in which case tree-level unitarity violation is due to longitudinal gauge boson scattering from the physical Higgs scalar \cite{b1}.

    Therefore if tree-level unitarity violation is an indication of true unitarity violation, then it is not possible to couple the Higgs doublet non-minimally to gravity as in \eq{e0}. New terms must also be added to \eq{e0}, in order to ensure unitarity is conserved at least up to energies sufficiently large compared with the value of $h$ during inflation, $h \approx \sqrt{N} M_{p}/\sqrt{\xi}$, where $N$ is the number of e-foldings of inflation.

\section{ A Unitarity-conserving completion of Higgs Inflation}

        As emphasized in \cite{nat}, the Einstein frame provides a particularly clear way to understand unitarity violation in graviton-mediated Higgs scattering due to the non-minimal coupling. On transforming to the Einstein frame, where the non-minimal couplings are eliminated, unitarity violation manifests itself via non-renormalizable interactions. Therefore the minimal unitarity-conserving completion of the Higgs Inflation Lagrangian in the Jordan frame will correspond to the Einstein frame Lagrangian which removes all the dangerous non-renormalizable terms.

     From the discussion of Section II, it is clear that the only way to eliminate unitarity violation in the Einstein frame is to replace the non-minimal Higgs kinetic term with a canonical kinetic term. We must therefore add terms to the Jordan frame action \eq{e4} to achieve this. The final action in the Einstein frame must have the form
 \be{e9}  S_{E} = \int d^{4} x  \sqrt{-\tilde{g}} \left( - \frac{M_{p}^{2}}{2}\tilde{R} + \tilde{g}^{\mu\nu}\left(D_{\mu}H\right)^{\dagger} \left(D_{\nu}H\right)
-\frac{1}{4} F_{\mu\nu} F^{\mu\nu}  - \frac{V(|H|)}{\Omega^4}    \right) ~.\ee
On transforming back to the Jordan frame, the additional terms in $S_{J}$ which are required to conserve unitarity up to the Planck scale are generated. The resulting unitarity-conserving action in the Jordan frame is given by
\bea
\label{e12}
S_J &=&  \int d^4\! x  \sqrt{-\!g} \left( - \frac{M_{p}^2 R}{2} - \xi H^{\dagger}HR + g^{\mu\nu}D_\mu H^{\dagger}D_\nu H
+ \frac{2 \xi H^{\dagger}H}{M_p^2} g^{\mu\nu}D_\mu H ^{\dagger}D_\nu H\right. \nonumber \\
& & \left. - \frac{3\xi^2}{\Omega^2M_p^2}g^{\mu\nu}\partial_\mu \left(H^{\dagger}H\right)\partial_\nu \left(H^{\dagger}H\right)
  -\frac{1}{4} F_{\mu\nu} F^{\mu\nu} - V(|H|) \right) ~.\eea
We believe that \eq{e12} is the minimal unitarity-conserving action for the Standard Model Higgs doublet with a large non-minimal coupling to gravity. Since the fundamental assumption of Higgs Inflation is that inflation is due entirely to the non-minimal coupling of $H^{\dagger}H$ to gravity, \eq{e12} will provide a manifestly unitarity-conserving basis for Higgs Inflation.

         The non-minimal coupling to $R$ plus the additional terms in \eq{e12} may be interpreted as the complete set of terms which must be brought down from the full Planck-scale gravity theory to the scale $\Lambda$ in order to maintain the quantum consistency of the theory.
A non-minimal coupling of the Higgs to gravity is generally expected to exist, but it is usually assumed that $\xi \sim 1$, in which case the associated unitarity violation occurs at $E \sim M_{p}$. The effect of increasing $\xi$ is to effectively pull down the non-minimal coupling from the Planck-scale gravity theory to the lower mass scale $\Lambda$. Unitarity violation can then be interpreted as a sign that other terms from the full gravity theory must accompany the non-minimal coupling in order to maintain the consistency of the theory.

    So far we have considered the model only at tree-level, without quantum corrections to the inflaton potential. The structure of \eq{e9} is equivalent to the Standard Model gauge and Higgs fields plus a potential $V(|H|)/\Omega^4$. This suggests that the 1-loop Coleman-Weinberg correction due to gauge boson loops in the Einstein frame will have the form $\sim M_{W}^4 \; \log \; M_{W}^2 \propto |H|^4$, which would spoil the flatness of the potential. In this case a supersymmetric (SUSY) version of the model will be necessary in order to suppress the quantum corrections to the inflaton potential. However, if the inflaton was not the Higgs, but instead a singlet scalar coupled to the Standard Model only via the potential (such as in \cite{rl1}), then its couplings would be suppressed by $\Omega^{-4}$ in the Einstein frame and should not spoil the flatness of the inflationary potential.

\section{Slow-roll inflation predictions}

  Although \eq{e12} provides a basis for a unitarity-conserving Higgs Inflation model, it is not the same Higgs Inflation model as originally proposed in \cite{bs1}. Inflation is best analysed in the Einstein frame, where $H$ has canonical kinetic terms and model may be treated as a conventional slow-roll inflation model, but now with potential $U(|H|) \equiv V(|H|)/\Omega^4$. Introducing the physical Higgs field as the inflaton, $H \rightarrow h/\sqrt{2}$, we obtain
\be{e12a} U(h) = \frac{ \lambda h^4}{4 \left(1 + \frac{\xi h^2}{M_{p}^2}\right)^2 } ~.\ee
For $h \gg M_{p}/\sqrt{\xi}$, the potential is flat and slow-roll inflation is possible. With $\tdN = 58$, where $\tilde{N} \approx \frac{\xi h^4}{16M_p^4}$ is the number of e-folding of inflation in the Einstein frame  (corresponding to $N = 60$ in the Jordan frame \cite{nat}), the classical value of the spectral index is given by $n = 1+2\eta - 6\epsilon$, where
\be{12b}  \eta \equiv M_p^2\left(\frac{d^2U}{dh^2}\right)  \simeq - \frac{12 M_p^4}{\xi h^4}  + \frac{36 M_p^6}{\xi^2 h^6};~\;\;\;\;
\epsilon \equiv \frac{M_p^2}{2} \left(\frac{1}{U}\frac{dU}{d h}\right)^2 \simeq \frac{8M_p^6}{\xi^2 h^6} - \frac{16 M_p^8}{\xi^3 h^8}
~.\ee
Therefore
\be{e13}   n \approx 1 - \frac{3}{2\tdN}  + \frac{3}{8\tdN^{3/2}\sqrt{\xi}}  \approx 0.974  ~.\ee
The tensor to scalar ratio $r$ is given by
\be{13b}
r \equiv 16\epsilon \simeq \frac{2}{\sqrt{\xi}\tdN^{3/2}} \approx 6 \times 10^{-6}
~.\ee
(The running of the spectral index $\alpha$ is negligibly small.)
The curvature perturbation is given by
\be{e13c}  P_{\xi} = \frac{ \lambda \tdN^3}{12 \pi^2 \xi^{3/2}}  ~,\ee
therefore to have a correctly normalized spectrum of density perturbations, $P_{\xi}^{1/2} = 4.8 \times 10^{-5}$, we require
\be{e14a}
\xi \simeq (3.8-6.5) \times 10^{5}~\ee
for $m_{H}$ in the range 114-170 GeV. This is different from the original Higgs Inflation model because the derivatives in the slow roll parameters are defined with respect to different canonically normalised fields - $\chi$ in the original model and $h$ in the unitarity-conserving model.
These may be compared with the predictions of the original Higgs Inflation model, $n \simeq 1 - \frac{2}{\tdN} - \frac{3}{2\tdN^2} = 0.965$, $r \simeq \frac{12}{\tdN^2} = 3.6 \times 10^{-3}$ and $ \frac{\lambda}{\xi^2} \simeq \frac{3(0.027)^4}{\tdN^2}$ giving $\xi \simeq 10^{4}$. 
These estimates are based on $N =60$.
It should be noted that, because the model contains only standard model parameters, it is in principle possible to determine the reheating temperature and hence $N$ precisely. Therefore the model has no free parameters.

\section{Conclusions}

      The possibility that inflation can be explained by a simple non-minimal coupling of the Higgs to gravity is very attractive, leading to a highly predictive model which requires no new fields beyond those of the Standard Model.  We have proposed a new Higgs Inflation model based on a unitarity-conserving extension of the original Higgs Inflation action. We believe that this is the minimal form of Higgs Inflation model which manifestly conserves unitarity in the presence of a non-minimal coupling of the Higgs to gravity. As such, it may provide the correct formulation of Higgs Inflation should strong coupling effects fail to eliminate unitarity violation in the original Higgs Inflation model.

     Perhaps the most interesting conclusion is that while unitarity-conserving Higgs Inflation is possible, the predictions of the new unitarity-conserving model are quite different from those of the original Higgs Inflation model. In particular, the classical spectral index of the new model is $n = 0.974$, which is within the 7-year WMAP 1-$\sigma$ limits on $n$ ($n = 0.963 \pm 0.012$ \cite{wmap7}) but significantly different from the original Higgs Inflation model prediction of $n = 0.965$. Therefore it should be possible to observationally distinguish between unitarity-conserving Higgs Inflation and the original Higgs Inflation model.

 We finally comment on the assumptions underlying our model. We
consider all terms which are scaled by inverse powers of $\Omega$ in the Einstein frame to lead to unitarity violation, with the exception of $V(|H|)/\Omega^{4}$. Terms are then added to eliminate unitarity violation.
However, we believe that $V(|H|)/\Omega^{4}$ will not lead to unitarity violation. This is  because in the limit $|H|^2 \gg M_{p}^{2}/2 \xi$, there is a nearly perfect cancellation of the $|H|^4$ factors in $V(|H|)$ and in $\Omega^4$, completely eliminating interactions\footnote{More generally, we expect that any non-polynomial potential interpolating between renormalizable potentials at small and large field strength will not lead to unitarity violation.}. To illustrate how the potential term differs from other terms with respect to unitarity violation, we can consider perturbations about a large background Higgs field. In this case the potential term in \eq{e4} tends towards that for massless, non-interacting scalars, with unitarity-violating interactions suppressed by powers of $|H|$, whereas the second term in \eq{e4}, for example,  leads to unitarity violation at $E \sim M_{p}/\sqrt{\xi}$, independent of $|H|$.

    A feature that the unitarity-conserving model shares with the original Higgs Inflation model is that since all the model parameters are Standard Model parameters, they can be fixed experimentally (with the exception of $\xi$, which is fixed by the density perturbations). In particular, it will be possible to precisely compute quantum corrections to the spectral index as a function of Higgs mass. This should allow for precision tests of the model once $m_{H}$ is determined by the LHC and $n$ by PLANCK. A caveat is that such quantum corrections are likely to be large in the case of a non-SUSY model, in which case a SUSY version following the same strategy will be necessary in order to maintain the flatness of the inflaton potential. A very minimal non-SUSY model may still be possible if the inflaton was instead a singlet scalar with a potential coupling to the Standard Model. However, we expect that the tree-level predictions of any unitarity-conserving model, being necessarily based on minimal kinetic terms and $V/\Omega^4$ in the Einstein frame, will remain unchanged.

\section*{Acknowledgements}

This work was supported by the European Union through the Marie Curie Research and Training Network "UniverseNet" (MRTN-CT-2006-035863).

\end{document}